\documentclass[journal]{IEEEtran}%
\usepackage{amsfonts}
\usepackage{amsmath}
\usepackage{amssymb}
\usepackage{graphicx}%
\providecommand{\U}[1]{\protect\rule{.1in}{.1in}}
\begin{document}
%
%TCIMACRO{\TeXButton{Title}{\title
%{Phenomenological Model for Predicting the Energy Resolution of Neutron-Damaged Coaxial HPGe Detectors}%
%}}%
%BeginExpansion
\title
{Phenomenological Model for Predicting the Energy Resolution of Neutron-Damaged Coaxial HPGe Detectors}%
%EndExpansion
%

%TCIMACRO{\TeXButton{Author}{\author
%{C. DeW. Van Siclen, E. H. Seabury, C. J. Wharton, and A. J. Caffrey
%\thanks{
%Manuscript received 2011. This work was supported by the Defense Threat Reduction Agency and the U. S. Department of Energy. It was performed at the Idaho National Laboratory, a DOE laboratory operated by Battelle Energy Alliance under DOE Idaho Operations Office Contract DE-AC07-05ID14517.}
%\thanks{
%The authors are with the Nuclear Nonproliferation Division, Idaho National Laboratory, P. O. Box 1625, Idaho Falls, ID 83415 USA. E-mail addresses: clinton.vansiclen@inl.gov, edward.seabury@inl.gov, jayson.wharton@inl.gov, gus.caffrey@inl.gov.}%
%}}}%
%BeginExpansion
\author{C. DeW. Van Siclen, E. H. Seabury, C. J. Wharton, and A. J. Caffrey
\thanks{
Manuscript received 2011. This work was supported by the Defense Threat Reduction Agency and the U. S. Department of Energy. It was performed at the Idaho National Laboratory, a DOE laboratory operated by Battelle Energy Alliance under DOE Idaho Operations Office Contract DE-AC07-05ID14517.}
\thanks{
The authors are with the Nuclear Nonproliferation Division, Idaho National Laboratory, P. O. Box 1625, Idaho Falls, ID 83415 USA. E-mail addresses: clinton.vansiclen@inl.gov, edward.seabury@inl.gov, jayson.wharton@inl.gov, gus.caffrey@inl.gov.}%
}%
%EndExpansion
%

%TCIMACRO{\TeXButton{pubid and specialpapernotice}{\IEEEpubid{}
%\IEEEspecialpapernotice{}}}%
%BeginExpansion
\IEEEpubid{}
\IEEEspecialpapernotice{}%
%EndExpansion
%

%TCIMACRO{\TeXButton{MakeTitle}{\maketitle}}%
%BeginExpansion
\maketitle
%EndExpansion
%

%TCIMACRO{\TeXButton{Begin Abstract}{\begin{abstract}
%}}%
%BeginExpansion
\begin{abstract}
%EndExpansion

The peak energy resolution of germanium detectors deteriorates with increasing
neutron fluence. This is due to hole capture at neutron-created defects in the
crystal which prevents the full energy of the gamma-ray from being recorded by
the detector. A phenomenological model of coaxial HPGe detectors is developed
that relies on a single, dimensionless parameter that is related to the
probability for immediate trapping of a mobile hole in the damaged crystal. As
this trap parameter is independent of detector dimensions and type, the model
is useful for predicting energy resolution as a function of neutron fluence.%

%TCIMACRO{\TeXButton{End Abstract}{\end{abstract}
%}}%
%BeginExpansion
\end{abstract}
%EndExpansion
%

%TCIMACRO{\TeXButton{Begin Key Words}{\begin{keywords}
%}}%
%BeginExpansion
%EndExpansion

\section{Introduction}

A gamma-ray traversing a germanium crystal loses energy mainly by the
production of Compton electrons (or a photoelectron in the case of low-energy
gamma-rays), which in turn lose energy by the production of electron-hole
pairs at roughly 2.96 eV each. The electrons and holes are collected at the
outer and inner electrodes covering the annular surfaces of the cylindrical
crystal, thereby recording the energy of the gamma-ray. The full energy is not
recovered, however, when holes are trapped at negatively charged defects in
the crystal created by a flux of fast neutrons. Thus the energy resolution of
gamma-ray peaks deteriorates with fast neutron fluence.

In practice, the gamma-ray energy is determined by measuring the current flow,
induced by the \textit{moving} electron and hole charges, between the
electrodes. In this phenomenological model, the gamma-ray energy is instead
calculated from the integrated charge at the two electrodes, induced by the
electron-hole pairs.

This paper is laid out as follows. In section II the relation between an
electron-hole pair and the induced charges at the electrodes is derived: this
is essentially the concept of the detector. In section III expressions for the
spatial density of hole traps (regions of fast neutron damage), and for the
interaction cross-section of those traps, are derived. In anticipation of the
computer implementation of the model, section IV describes the stochastic
methods for choosing the location of electron-hole pair creation and
subsequent location of the hole trapping. In section V the computer program,
that combines these pieces of the model to produce spectral line peaks, is
described. Finally, model results are compared to measurements, made in our
laboratory, of the energy resolution of the 1332 keV $^{60}$Co spectral line
obtained by a p-type detector irradiated by a fast neutron fluence up to
10$^{9}$ cm$^{-2}$.

The primary purpose of this work is to produce a predictive model of peak
energy resolution as a function of fast neutron fluence. It is a
phenomenological model, meaning that critical parameters are taken from
experiment rather than calculated from first principles. The (dimensionless)
critical parameters in this case are $\alpha_{h}$, which is related to the
hole trap cross-section, and $\alpha_{t}$, which is essentially the average
number of hole traps created by a fast neutron as it collides with atoms in
the germanium crystal. (These are both introduced in section III.) Values for
these parameters are obtained by reproducing, with the model, the data of R.
H. Pehl \textit{et al.} \cite{r1}. In that work, two HPGe coaxial detectors
(one n-type, the other p-type), fabricated from the same crystal, were
irradiated simultaneously with fast neutrons from an unmoderated $^{252} $Cf source.

For the convenience of the reader, SI units (and derived units) are used in
any calculations: electric potential $\phi$ in volts (V); electric field $E$
in V$\cdot$m$^{-1}$; force in newtons (n); charge in coulombs (C). Note that 1
V = 1 n$\cdot$m$\cdot$C$^{-1}$. Further, the product $e$ ($v$ V) = $v$ eV. The
electric charge $e=1.602\times10^{-19}$ C; the vacuum permittivity
$\epsilon_{0}=8.854\times10^{-12}$ C$^{2}\cdot$n$^{-1}\cdot$m$^{-2}$.

Where the symbols $\pm$ and $\mp$ occur, the top/bottom sign is used for
p-type/n-type detectors. The coaxial detector is an annular cylinder so it is
natural to use cylindrical coordinates $\left(  r,z\right)  $: the radius of
the inner electrode is $R_{0}$; the radius of the outer electrode is $R_{1}$;
the axial coordinate $z=0$ at the top of the coaxial detector (the end
pointing at the radiation source), and $z>0$ within the detector.

\section{Induced charge at the electrodes due to electron-hole pairs}

As an electron and hole have opposite charge, it is not until they move apart
under the influence of the driving force $q\mathbf{E}$($r$) that charge is
induced at the electrodes. The variable $q$ is the charge of the particle (so
equals $+e$/$-e$ for holes/electrons) and $\mathbf{E}(r)$ is the electric
field at the particle location $r$ between the electrodes. In the case of
p-type detectors, the outer contact is positively biased so that the field
$\mathbf{E}(r)$ is negative; thus holes move towards the inner contact at
$R_{0}$ and electrons move towards the outer contact at $R_{1}$. In the case
of n-type detectors, the inner contact is positively biased so that the field
$\mathbf{E}(r)$ is positive; thus holes/electrons move towards the outer/inner contact.

What charges are induced at the electrodes by an electron-hole pair after
separation? Expressions for these are easily found by use of Green's
reciprocation theorem \cite{r1b}: For a given arrangement of electrodes, if
$\phi$ is the potential due to a volume-charge distribution $\rho$ and a
surface-charge distribution $\sigma$, while $\phi^{\prime}$ is the potential
due to other charge distributions $\rho^{\prime}$ and $\sigma^{\prime}$, then%
\begin{equation}
\int\nolimits_{V}\rho\phi^{\prime}d^{3}x+\int\nolimits_{S}\sigma\phi^{\prime
}da=\int\nolimits_{V}\rho^{\prime}\phi d^{3}x+\int\nolimits_{S}\sigma^{\prime
}\phi da\text{.} \label{e1}%
\end{equation}
In the \textquotedblleft primed\textquotedblright\ system, set $\rho^{\prime
}=\sigma^{\prime}=0$, so $\phi^{\prime}(r)$ is the solution to Laplace's
equation in cylindrical coordinates with boundary conditions $\phi^{\prime}=1$
at surface $S$, which is inner contact $R_{0}$, and $\phi^{\prime}=0$ at outer
contact $R_{1}$. Then the right hand side equals zero, so the
\textquotedblleft unprimed\textquotedblright\ system (which is the annular
cylinder with a point charge at $r$ and a surface charge at the inner contact)
obeys the relation
\begin{equation}
\int\nolimits_{V}\rho\phi^{\prime}d^{3}x=-\int\nolimits_{S}\sigma da\text{.}
\label{e2}%
\end{equation}
Recognizing that $\rho$ (the charged particle at $r$) is $q$ times the delta
function produces the relation $q\phi^{\prime}(r)=-\int\nolimits_{S}\sigma
da$. Thus the induced charge at the \textit{inner} contact due to a charged
particle at $r$ is%
\begin{align}
Q_{0}(r)  &  =-q\phi^{\prime}(r)\nonumber\\
&  =-q\left(  1-\frac{\ln\left[  r/R_{0}\right]  }{\ln\left[  R_{1}%
/R_{0}\right]  }\right)  =-q\frac{\ln\left[  R_{1}/r\right]  }{\ln\left[
R_{1}/R_{0}\right]  }\text{.} \label{e3}%
\end{align}
Similarly (but setting $\phi^{\prime}=1$ at surface $S$, which is outer
contact $R_{1}$, and $\phi^{\prime}=0$ at inner contact $R_{0}$), the induced
charge $Q_{1}(r)$ at the \textit{outer} contact due to a charged particle at
$r$ is%
\begin{equation}
Q_{1}(r)=-q\phi^{\prime}(r)=-q\frac{\ln\left[  r/R_{0}\right]  }{\ln\left[
R_{1}/R_{0}\right]  }\text{.} \label{e4}%
\end{equation}
Note that the induced charges $Q_{0}(r)$ and $Q_{1}(r)$ at the two electrodes
are opposite in sign to the charge $q$ of the particle at $r$, and that the
sum $Q_{0}(r)+Q_{1}(r)$ always equals $-q$.

Now consider an electron-hole pair created at $r_{i}$. As the two particles
are of opposite sign and both reside at $r_{i}$, no charge is induced at the
two contacts. However, as the two mobile charges separate and move radially to
$r_{e}$ and $r_{h}$, respectively, the charge $Q_{0}$ is induced at the
$R_{0}$ contact and the charge $Q_{1}$ is induced at the $R_{1}$ contact,
where%
\begin{equation}
Q_{0}=Q_{0}(r_{e})+Q_{0}(r_{h})=e\frac{\ln\left[  r_{h}/r_{e}\right]  }%
{\ln\left[  R_{1}/R_{0}\right]  } \label{e5}%
\end{equation}%
\begin{equation}
Q_{1}=Q_{1}(r_{e})+Q_{1}(r_{h})=-e\frac{\ln\left[  r_{h}/r_{e}\right]  }%
{\ln\left[  R_{1}/R_{0}\right]  }\text{.} \label{e6}%
\end{equation}
It is these induced charges $Q_{0}$ and $Q_{1}$ that account for the current
between the two contacts due to an electron-hole pair (note that an
incremental change $\delta Q_{0}=-\delta Q_{1}$ always, as electric charge
flows from one contact to the other). Note that initially $r_{e}=r_{h}=r_{i}$
so the induced charges $Q_{0}=Q_{1}=0$, and that when both mobile charges
successfully reach their respective contacts, the induced charges $Q_{0}=\mp
e$ and $Q_{1}=\pm e$, as expected (top/bottom sign indicates p-type/n-type
detector). As the current between electrodes is just the transfer of charge
from one to the other, the result $\left\vert Q_{0}\right\vert =\left\vert
Q_{1}\right\vert =e$ allows the full creation energy of the electron-hole pair
to be recorded.

For a p-type detector, when the hole is trapped at $r_{h}$ but the electron
successfully reaches $R_{1}$, the charge induced at the central contact is
$Q_{0}=-e\frac{\ln\left[  R_{1}/r_{h}\right]  }{\ln\left[  R_{1}/R_{0}\right]
}$ and the charge induced at the outer contact is $Q_{1}=e\frac{\ln\left[
R_{1}/r_{h}\right]  }{\ln\left[  R_{1}/R_{0}\right]  }$. Note that in this
case $Q_{0}=Q_{0}\left(  r_{h}\right)  $ and $Q_{1}=e+Q_{1}\left(
r_{h}\right)  <e$, meaning that $\left\vert Q_{0}\right\vert =\left\vert
Q_{1}\right\vert <e$ so not all of the pair creation energy is recorded. For
an n-type detector, when the hole is trapped at $r_{h}$ but the electron
successfully reaches $R_{0}$, the induced charges are $Q_{0}=e\frac{\ln\left[
r_{h}/R_{0}\right]  }{\ln\left[  R_{1}/R_{0}\right]  }$ and $Q_{1}=-e\frac
{\ln\left[  r_{h}/R_{0}\right]  }{\ln\left[  R_{1}/R_{0}\right]  }$, and again
not all of the pair creation energy is recorded. By examining the two
expressions for $Q_{0}$ (or $Q_{1}$), it is evident that in general the
average value $\left\langle \left\vert Q_{0}\right\vert \right\rangle $ for a
damaged p-type detector will be less than that for a damaged n-type detector,
since a spatially uniform flux of gamma-rays will produce more electron-hole
pairs near the outer contact than near the inner contact, so resulting in more
hole trapping near the outer contact (thus $\left\langle \ln\left[
R_{1}/r_{h}\right]  \right\rangle <\left\langle \ln\left[  r_{h}/R_{0}\right]
\right\rangle $). This effect translates into less energy being attributed to
an incident gamma-ray by a p-type detector than by an n-type detector.

The induced charges (currents) at the contacts are related to the energy of
the incident gamma-ray in a straightforward way. When the electron and hole
reach their terminal locations at $r_{e}$ and $r_{h}$, respectively, the
induced charges $Q_{0}$ and $Q_{1}$ correspond to an energy $\frac{Q}%
{e}\epsilon$ recorded by the detector, where $Q\equiv\left\vert Q_{0}%
\right\vert =\left\vert Q_{1}\right\vert $ and $\epsilon$ is the average
energy needed to create an electron-hole pair (this is the energy needed to
elevate an electron in the valence band into the conduction band). For
germanium, $\epsilon=2.96$ eV at 77 K \cite{r1e}. An incident gamma-ray of
energy $E_{\gamma}$ that produces $n$ electron-hole pairs will thus record an
energy $n\frac{\left\langle Q\right\rangle }{e}\epsilon$, where the average
value $\left\langle Q\right\rangle $ is taken over all the $n$ pairs.

In the implementation of this model in a computer code, a value for $n$ is
chosen for each incident gamma-ray from the Gaussian distribution
$p(n)=\left(  2\pi\sigma^{2}\right)  ^{-1/2}\exp\left[  -\left(  n-n_{\gamma
}\right)  ^{2}/\left(  2\sigma^{2}\right)  \right]  $. The average value
$n_{\gamma}=E_{\gamma}/\epsilon$ [so for example, the $1332$ keV $^{60}$Co
gamma-ray produces (on average) $n_{\gamma}=4.5\times10^{5}$ electron-hole
pairs in a germanium detector]. The variance $\sigma^{2}=F\cdot n_{\gamma}$
where $F$ is the Fano factor (and is approximately $0.13$ for Ge detectors
\cite{r1e}). To obtain values for $n$ from this distribution, it is most
convenient to use the Box-Muller method \cite{r2}: Consider the
\textquotedblleft standard normal\textquotedblright\ distribution $\phi
(\nu)=\left(  2\pi\right)  ^{-1/2}\exp\left[  -\nu^{2}/2\right]  $. Then the
two random variables $\nu_{1}$ and $\nu_{2}$ will both have the standard
normal distribution and will be independent, where%
\begin{equation}
\nu_{1}=\left(  -2\ln x_{1}\right)  ^{1/2}\cos\left(  2\pi x_{2}\right)
\label{e7}%
\end{equation}%
\begin{equation}
\nu_{2}=\left(  -2\ln x_{1}\right)  ^{1/2}\sin\left(  2\pi x_{2}\right)
\label{e8}%
\end{equation}
and $x_{1}$ and $x_{2}$ are random numbers taken from the uniform distribution
on $(0,1]$. Then the desired value $n=n_{\gamma}+\sigma\nu$, where $\nu$ is
either $\nu_{1}$ or $\nu_{2}$ calculated from Eq. (\ref{e7}) or Eq. (\ref{e8}).

Note that the FWHM of the Gaussian distribution $p(n)$ is $2\left(
2\sigma^{2}\ln2\right)  ^{1/2}$. The FWHM of the corresponding energy peak is
then $2\left(  2Fn_{\gamma}\ln2\right)  ^{1/2}\cdot\epsilon=2\left(
2FE_{\gamma}\epsilon\ln2\right)  ^{1/2}$. So for example, in the absence of
neutron damage, the FWHM of the peak corresponding to the $1332$ keV $^{60}$Co
spectral line is $1.686$ keV, while that of the peak corresponding to the
$122.1$ keV $^{57}$Co spectral line is $0.510$ keV.

\section{Hole trapping}

The fate of mobile holes (and electrons) in the neutron-damaged crystal is
determined by the spatial distribution of traps, and the trap cross-section.
These two functions are derived in turn.

The distribution of particle traps should be uniform over a z-slice (thickness
$dz$), since the source of gamma-rays (at which the detector is pointed) also
acts as the primary source of neutrons. What then is the trap density
$\rho_{q}(z)$ (the number of traps in the infinitesimal volume $A\cdot dz$ at
axial position $z$, where $A$ is the cross-sectional area of the crystal)?
Consider that the mean free path of an incident neutron is $l$. That is, the
probability that the neutron first collides a distance $\left[  z,z+dz\right]
$ from its entry point into the crystal is $\exp\left[  -z/l\right]
\cdot\left(  dz/l\right)  $. Then $N\exp\left[  -z/l\right]  \cdot\left(
dz/l\right)  $ is the number of neutrons that first collide a distance
$\left[  z,z+dz\right]  $ from their entry point at $z=0$, where $N$ is the
number of neutrons incident on the crystal. Assuming that the collision
produces a trap, the trap density%
\begin{align}
\rho_{q}(z)  &  =\alpha_{t}\cdot N\exp\left[  -z/l\right]  \cdot\left(
dz/l\right)  /\left(  A\cdot dz\right) \nonumber\\
&  =\alpha_{t}\cdot\frac{N}{lA}\exp\left[  -z/l\right]  \label{e9}%
\end{align}
where $N/A$ is the neutron fluence, and the dimensionless parameter
$\alpha_{t}$ is the (average) number of traps created by a fast neutron as it
collides with atoms in the germanium crystal. Note that $\alpha_{t}$\ should
be somewhat larger than $1$, to account---in a crude way---for any subsequent
collisions by the neutron.

What is a reasonable value for $\alpha_{t}$? For a neutron incident on a
detector of length (thickness) $L$, the probability that \textit{no}
collisions occur over the distance $L$ is $\exp\left[  -L/l\right]  $, meaning
that for a neutron fluence $N/A$, the fraction $1-\exp\left[  -L/l\right]  $
undergo at least one collision. L. S. Darken \textit{et al.} \cite{r3}, after
irradiating a $3$ cm thick Ge crystal with neutrons, conclude
\textquotedblleft A fast neutron flux of $10^{10}$ cm$^{-2}$ produces about
$2\times10^{9}$ cm$^{-3}$ disordered regions of various
sizes\textquotedblright. Setting $l=6$ cm \cite{r4} and $L=3$ cm, the fraction
of incident neutrons that underwent collisions in the crystal was $0.3935$.
Each of those neutrons was then responsible for $\left(  2\times10^{9}%
\times3\right)  /\left(  0.3935\times10^{10}\right)  =1.5$ traps per
(collided) neutrons. Thus in general $\alpha_{t}\sim1.5$.

As charged particles, holes and electrons may be trapped at defects with
opposite charge. It is believed that hole traps are large disordered regions
with large negative charge \cite{r4e}, and so have a large effective
cross-section, while electron traps are much smaller (perhaps point defects)
and so have a much smaller cross-section. In any event, the trap cross-section
$\sigma_{q}$ must be roughly the size of the local distortion, due to the
electric charge of the defect, of the applied electric field $\mathbf{E}(r)$.
An expression for $\sigma_{q}$ can be derived as follows.

For simplicity, use 2D Cartesian coordinates, and place the defect (with
charge $Q$) at the origin, and set the no-defect electric field $\mathbf{E}%
=E\widehat{\mathbf{x}}$. Then the potential $\phi^{\ast}(\mathbf{r})$ at the
point $\mathbf{r}=(x,y)$ for this system is $\phi\left(  x\right)  +\frac
{Q}{\epsilon}\left(  x^{2}+y^{2}\right)  ^{-1/2}$ where $\phi\left(  x\right)
$ is the potential at $\mathbf{r}$ in the absence of the defect and $\epsilon$
is the permittivity of germanium. This produces the electric field%
\begin{align}
\mathbf{E}^{\ast}\left(  \mathbf{r}\right)   &  =-\mathbf{\nabla}\phi^{\ast
}\left(  \mathbf{r}\right)  =\left\{  E+\frac{Q}{\epsilon}x\left(  x^{2}%
+y^{2}\right)  ^{-3/2}\right\}  \widehat{\mathbf{x}}\nonumber\\
&  +\frac{Q}{\epsilon}y\left(  x^{2}+y^{2}\right)  ^{-3/2}\widehat{\mathbf{y}%
}\text{.} \label{e10}%
\end{align}
The electric field lines in the absence of the defect are directed parallel to
the x-axis; in the presence of the charged defect at the origin, they are bent
towards the origin. Those field lines that terminate at the defect are
particle paths that lead to trapping. Clearly the field lines are seriously
bent towards the defect when the magnitude of the y-component $\frac
{Q}{\epsilon}y\left(  x^{2}+y^{2}\right)  ^{-3/2}$ of the field $\mathbf{E}%
^{\ast}\left(  \mathbf{r}\right)  $\ exceeds that of the x-component
$E+\frac{Q}{\epsilon}x\left(  x^{2}+y^{2}\right)  ^{-3/2}$; that is, when
$x\approx0$ and $y^{2}<\left\vert \frac{Q}{\epsilon E}\right\vert $. Thus the
trap cross-section $\sigma\sim\pi y^{2}\propto\frac{e}{\epsilon E}$. As the
electric field $\mathbf{E}(r)$ in the crystal has a radial dependence,
$\sigma_{q}(r)=\alpha_{q}\cdot\frac{e}{\epsilon}E(r)^{-1}$ where $\alpha_{q}$
is a dimensionless parameter.

What are reasonable values for $\alpha_{q}$? L. S. Darken \textit{et al.}
\cite{r5} estimate cross-sections $\sigma_{h}\sim10^{-11}$ cm$^{2}$ and
$\sigma_{e}\sim10^{-13}$ cm$^{2}$. A typical value for $E(r)$ (the magnitude
of the electric field in the crystal produced by the bias potential at an
electrode) is 125 kV m$^{-1}$. Thus $\alpha_{h}\sim0.1$ and $\alpha_{e}%
\sim10^{-3}$.

Due to the higher production of electron-hole pairs near the outer contact
$R_{1}$, it is preferable to have a larger electric field $E$ there as well to
reduce the trap cross-sections $\sigma_{q}$. This shaping of the electric
field is accomplished by doping p-type and n-type germanium detectors with
(electron acceptor) boron and (electron donor) lithium, respectively. These
dopants produce an intrinsic space (free) charge density $\rho=-e\rho
_{\text{A}}$ in the case of p-type detectors and $\rho=e\rho_{\text{D}}$ in
the case of n-type detectors \cite{r6}, where $\rho_{\text{A/D}}$ is the
density of acceptor/donors. A typical value is $\rho_{\text{A/D}}=10^{10}$
cm$^{-3}$.

The potential $\phi(r)$ between the contacts satisfies Poisson's equation,
$\nabla^{2}\phi=-\rho/\epsilon$, where $\epsilon=16\epsilon_{0}$ is the
permittivity of Ge. This equation%
\begin{equation}
\frac{1}{r}\frac{d}{dr}\left(  r\frac{d}{dr}\phi(r)\right)  =-\frac{\rho
(r)}{\epsilon} \label{e11}%
\end{equation}
is solved for $\phi(r)$ given the boundary conditions, which are the applied
potentials at the outer and inner contacts. For p-type/n-type detectors, the
outer/inner contact is positively biased. The electric field between the
contacts is then $\mathbf{E}(r)=-\mathbf{\nabla}\phi(r)$.

In the usual case that the charge density $\rho$ has no radial dependence,
$\phi(r)=\phi\left(  R_{0}\right)  -\frac{\rho}{4\epsilon}\left(  r^{2}%
-R_{0}^{2}\right)  +A\ln\left[  r/R_{0}\right]  $ and $\mathbf{E}(r)=\left(
\frac{\rho}{2\epsilon}r-\frac{A}{r}\right)  \widehat{\mathbf{r}}$ where the
constant%
\begin{equation}
A=\frac{\phi\left(  R_{1}\right)  -\phi\left(  R_{0}\right)  +\frac{\rho
}{4\epsilon}\left(  R_{1}^{2}-R_{0}^{2}\right)  }{\ln\left[  R_{1}%
/R_{0}\right]  }\text{.} \label{e12}%
\end{equation}
Thus the electric field magnitude $E(r)$, needed to calculate the trap
cross-sections $\sigma_{q}(r)$, is easily obtained.

\section{Trapping probability}

Trapping of a mobile, charged particle (electron or hole) is a stochastic
process, meaning that the probability that a particle at $r^{\prime}$ will be
trapped in the infinitesimal distance interval $\left[  r^{\prime},r^{\prime
}+dr^{\prime}\right]  $ is $\rho_{q}(z)\sigma_{q}(r^{\prime})dr^{\prime}$.
Then the probability that it will \textit{not} be immediately trapped is
$1-\rho_{q}(z)\sigma_{q}(r^{\prime})dr^{\prime}$, which effectively equals
$\exp\left[  -\rho_{q}(z)\sigma_{q}(r^{\prime})dr^{\prime}\right]  $. By
taking the product of many such exponentials, the probability $p_{q}$ for the
charged particle created at $r_{i}$ to successfully reach $r$ is%
\begin{equation}
p_{q}\left(  r_{i},r,z\right)  =\exp\left[  -\left\vert \int\limits_{r_{i}%
}^{r}\rho_{q}(z)\sigma_{q}(r^{\prime})dr^{\prime}\right\vert \right]
\label{e13}%
\end{equation}
where use of the absolute value allows for the case ($r<r_{i}$) that the
particle moves towards the inner electrode.

The probability for the particle, having been created at the interaction point
$r_{i}$, to be subsequently trapped in the infinitesimal interval $\left[
r,r+dr\right]  $ is then $-\frac{dp_{q}\left(  r_{i},r,z\right)  }{dr}dr$. To
see this, note that%
\begin{align}
-\int\limits_{r_{i}}^{r}\frac{dp_{q}\left(  r_{i},r^{\prime},z\right)
}{dr^{\prime}}dr^{\prime}  &  =-p_{q}\left(  r_{i},r,z\right)  +p_{q}\left(
r_{i},r_{i},z\right) \nonumber\\
&  =1-p_{q}\left(  r_{i},r,z\right)  \label{e14}%
\end{align}
is the probability that a particle, created at $r_{i}$, never arrives at $r$.
Thus the derivative $-\frac{dp_{q}\left(  r_{i},r,z\right)  }{dr}\equiv
T_{q}\left(  r_{i},r,z\right)  $ is the PDF (probability distribution
function) for the particle trap position $r$ given $r_{i}$.

To perform a computer simulation of particle creation and trapping, a trap
position $r$ is randomly selected from this distribution. How is this done?
The formula for converting a random number $x$ taken from the \textit{uniform}
probability distribution $P(x)=1$ (such $x$ values are produced by standard
random number generators) to the corresponding $r$ value is derived as
follows. The probabilities $T_{q}\left(  r_{i},r,z\right)  dr$ and $P(x)dx$
must be equal, so $T_{q}\left(  r_{i},r,z\right)  dr=dx$. Then integrating the
former from $r_{i}$ to $r$, and the latter from $0$ to $x$ gives%
\begin{align}
x  &  =\int\limits_{r_{i}}^{r}T_{q}\left(  r_{i},r^{\prime},z\right)
dr^{\prime}\nonumber\\
&  =-\int\limits_{r_{i}}^{r}\frac{dp_{q}\left(  r_{i},r^{\prime},z\right)
}{dr^{\prime}}dr^{\prime}=1-p_{q}\left(  r_{i},r,z\right)  \label{e15}%
\end{align}
which relates a randomly chosen $x$ value to an $r$ value. The function
$x(r)=1-p_{q}\left(  r_{i},r,z\right)  $ must be inverted so as to give $r$
when $x$ is chosen randomly from the interval $[0,1)$; that is, the function
$r(x)$ must be found. This is done by expressing the relation as%
\begin{align}
\ln\left[  1-x\right]   &  =\ln\left[  p_{q}\left(  r_{i},r,z\right)  \right]
=-\left\vert \int\limits_{r_{i}}^{r}\rho_{q}(z)\sigma_{q}(r^{\prime
})dr^{\prime}\right\vert \nonumber\\
&  =-\rho_{q}(z)\frac{\alpha_{q}e}{16\epsilon_{0}}\left\vert \int%
\limits_{r_{i}}^{r}\frac{dr^{\prime}}{E(r^{\prime})}\right\vert \text{.}
\label{e16}%
\end{align}
This integral can be solved analytically by noting that $E(r)^{-1}=\frac
{r}{-A+\frac{\rho}{2\epsilon}r^{2}}$, giving (after some careful algebra)%
\begin{equation}
\ln\left[  1-x\right]  =\pm\frac{q}{e}\rho_{q}(z)\frac{\alpha_{q}}%
{\rho_{\text{A/D}}}\ln\left[  \frac{-A+\frac{\rho}{2\epsilon}r^{2}}%
{-A+\frac{\rho}{2\epsilon}r_{i}^{2}}\right]  \label{e17}%
\end{equation}
under the condition that $\rho\neq0$ (i.e., that the crystal is doped).

If the $x$ value chosen from the interval $[0,1)$ is greater than
$1-p_{q}\left(  r_{i},R,z\right)  $, where $R$ is the radius of the contact to
which the particle is moving, then the particle has successfully reached that
contact. That is, if the $x$ value satisfies the relation%
\begin{equation}
\ln\left[  1-x\right]  \leq\pm\frac{q}{e}\rho_{q}(z)\frac{\alpha_{q}}%
{\rho_{\text{A/D}}}\ln\left[  \frac{-A+\frac{\rho}{2\epsilon}R^{2}}%
{-A+\frac{\rho}{2\epsilon}r_{i}^{2}}\right]  \label{e18}%
\end{equation}
then the particle has successfully reached the contact at radius $R$.
Otherwise the position $r$ at which the particle is trapped is obtained from
the equation%
\begin{equation}
r^{2}=\frac{2\epsilon}{\rho}\left\{  A+\left(  -A+\frac{\rho}{2\epsilon}%
r_{i}^{2}\right)  \left(  1-x\right)  ^{\pm\frac{q}{e}\frac{\rho_{\text{A/D}}%
}{\rho_{q}(z)\alpha_{q}}}\right\}  \text{.} \label{e19}%
\end{equation}

A similar stochastic approach is taken for obtaining $r_{i}$, the radial
location at which a gamma-ray enters the crystal. For simplicity, the
gamma-ray is assumed to shed\ $n$ electron-hole pairs at random points
$\left(  r_{i},z\right)  $ as it traverses the length of the crystal. Since
the z-axis of the detector points at the gamma-ray source, the areal
distribution of $r_{i}$ in a simulation should be uniform over a z-slice of
the crystal. Thus the areal density of $r_{i}$ points is constant: call it
$\rho_{i}$ (points per area). Then $dx=\rho_{i}dA$ giving%
\begin{equation}
x=\frac{%
%TCIMACRO{\tint \nolimits_{R_{0}}^{r_{i}}}%
%BeginExpansion
{\textstyle\int\nolimits_{R_{0}}^{r_{i}}}
%EndExpansion
\rho_{i}dA}{%
%TCIMACRO{\tint \nolimits_{R_{0}}^{R_{1}}}%
%BeginExpansion
{\textstyle\int\nolimits_{R_{0}}^{R_{1}}}
%EndExpansion
\rho_{i}dA}=\frac{%
%TCIMACRO{\tint \nolimits_{R_{0}}^{r_{i}}}%
%BeginExpansion
{\textstyle\int\nolimits_{R_{0}}^{r_{i}}}
%EndExpansion
\rho_{i}2\pi rdr}{%
%TCIMACRO{\tint \nolimits_{R_{0}}^{R_{1}}}%
%BeginExpansion
{\textstyle\int\nolimits_{R_{0}}^{R_{1}}}
%EndExpansion
\rho_{i}2\pi rdr}=\frac{r_{i}^{2}-R_{0}^{2}}{R_{1}^{2}-R_{0}^{2}}\text{.}
\label{e20}%
\end{equation}
Inverting the function $x\left(  r_{i}\right)  $ gives%
\begin{equation}
r_{i}=\left\{  R_{0}^{2}+x\left(  R_{1}^{2}-R_{0}^{2}\right)  \right\}  ^{1/2}
\label{e21}%
\end{equation}
from which a value $r_{i}$ is obtained by randomly choosing a value $x$ from
the interval $\left[  0,1\right]  $.

\section{Simulation algorithm}

The pieces developed above are assembled into a computer model of a coaxial
HPGe detector. The inputs are the parameter values for the detector
($R_{0},R_{1},L,\phi\left(  R_{0}\right)  ,\phi\left(  R_{1}\right)
,\rho_{\text{A/D}}$), the spectral line ($E_{\gamma}$) of interest, and the
neutron fluence $N/A$. Then the model considers the gamma-rays emitted from a
source to be normally incident on the $z=0$\ (top) surface of the detector;
subsequently each gamma-ray maintains its radial position $r_{i}$ and produces
$n$ electron-hole pairs as it traverses the length of the crystal. For each
gamma-ray, the values $r_{i}$ and $n$ are obtained stochastically according to
Eq. (\ref{e21}) and as described at the end of section II, respectively, and
the $n$ pair creation points $\left(  r_{i},z\right)  $ are distributed
randomly over the length $L$ (that is, the $z$ value for a pair is taken
randomly from the interval $\left[  0,L\right]  $ of the uniform distribution).

The contribution of each electron-hole pair to the recorded energy of the
gamma-ray is obtained by the following steps: (i) The trap density $\rho
_{q}(z)$ is calculated for each particle by Eq. (\ref{e9}). (ii) This allows
the terminal positions $r_{e}$ and $r_{h}$ to be obtained stochastically
according to Eqs. (\ref{e18}) and (\ref{e19}). (iii) Using these values
$r_{e}$ and $r_{h}$, the induced charge $Q_{0}$ at the inner electrode is
calculated by Eq. (\ref{e5}). (iv) Then the contribution of the electron-hole
pair to the recorded energy is $\left(  \left\vert Q_{0}\right\vert /e\right)
\times2.96$ eV. The contributions of all $n$ pairs constitute the recorded
energy of the gamma-ray.

The energy peak is constructed as a histogram of the gamma-ray energies. The
finite width of the peak is due to the variation in the $n$ value (around the
average value $n_{\gamma}$) for gamma-rays producing the peak, and to hole and
electron trapping suffered by some fraction of the $n$ pairs produced by each
gamma-ray. Thus the width of the peak can be modified by adjusting the values
of the dimensionless parameters $\alpha_{t}$ and $\alpha_{q}$, or rather, the
\textit{product} $\alpha_{t}\cdot\alpha_{q}\equiv A_{q}$. According to section
III, the value $A_{h}$ should lie in the range $\left[  0.1,1\right]  $, and
$A_{e}$ should be two orders of magnitude or so smaller.

\begin{figure}[ptb]%
\centering
\includegraphics[
height=2.7259in,
width=3.2915in
]%
{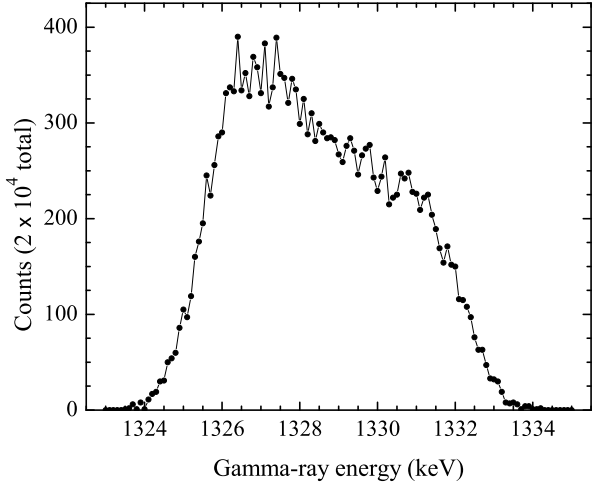}%
\caption{Peak corresponding to the 1332 keV $^{60}$Co spectral line obtained
by the p-type detector model for a neutron fluence of 10$^{9}$ cm$^{-2}$. This
model detector has the same dimensions and bias potential as the p-type
detector studied by Pehl \textit{et al.} \cite{r1}.}%
\label{fig1}%
\end{figure}

This \textquotedblleft tuning\textquotedblright\ is accomplished by
reproducing the data of R. H. Pehl \textit{et al.} \cite{r1}. As mentioned
above, two HPGe coaxial detectors (one n-type, the other p-type), fabricated
from the same crystal, were irradiated simultaneously with fast neutrons from
an unmoderated $^{252}$Cf source. The input to the model is the following:
inner radius $R_{0}=4$ mm; outer radius $R_{1}=21$ mm; crystal length $L=30$
mm; bias potential for the p-type detector $\phi\left(  R_{1}\right)  =1.6$
kV; bias potential for the n-type detector $\phi\left(  R_{0}\right)  =2.8$
kV. As dopant densities are not provided, the typical value $\rho_{\text{A/D}%
}=10^{10}$ cm$^{-3}$ is used. The key data point from Ref. \cite{r1} is the
FWHM resolution of $6$ keV for the 1332 keV $^{60}$Co line obtained by the
p-type detector after a neutron fluence of 10$^{9}$ cm$^{-2}$. This
experimental result is reproduced by the model when $A_{h}=0.30$ (and
$A_{e}=0.001$), as shown in Fig. \ref{fig1}.%
%TCIMACRO{\FRAME{ftbpFU}{3.2915in}{2.7259in}{0pt}{\Qcb{Peak corresponding to
%the 1332 keV $^{60}$Co spectral line obtained by the p-type detector model for
%a neutron fluence of 10$^{9}$ cm$^{-2}$. This model detector has the same
%dimensions and bias potential as the p-type detector studied by Pehl
%\QTR{it}{et al.} \cite{r1}.}}{\Qlb{fig1}}{vansi1.eps}%
%{\special{ language "Scientific Word";  type "GRAPHIC";
%maintain-aspect-ratio TRUE;  display "USEDEF";  valid_file "F";
%width 3.2915in;  height 2.7259in;  depth 0pt;  original-width 4.0534in;
%original-height 3.4238in;  cropleft "0";  croptop "1";  cropright "1";
%cropbottom "0";  filename 'vansi1.EPS';file-properties "XNPEU";}}}%
%BeginExpansion
%EndExpansion

\begin{figure}[ptb]%
\centering
\includegraphics[
height=2.7259in,
width=3.2915in
]%
{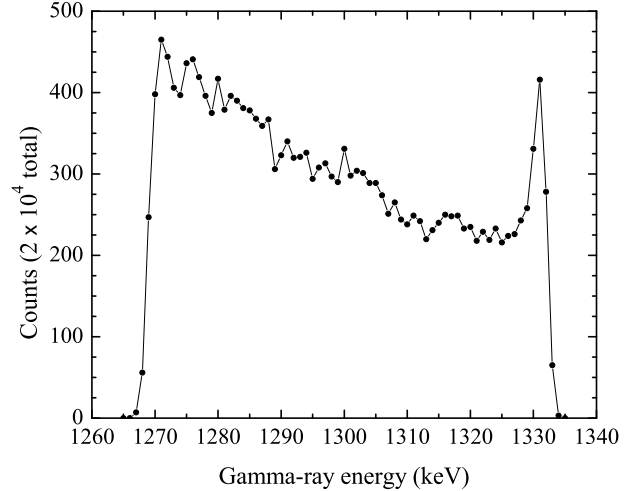}%
\caption{Peak corresponding to the 1332 keV $^{60}$Co spectral line obtained
by the p-type detector model for a neutron fluence of 10$^{10}$ cm$^{-2}$.
This model detector has the same dimensions and bias potential as the p-type
detector studied by Pehl \textit{et al.} \cite{r1}.}%
\label{fig2}%
\end{figure}

\begin{figure}[ptb]%
\centering
\includegraphics[
height=2.7371in,
width=3.2915in
]%
{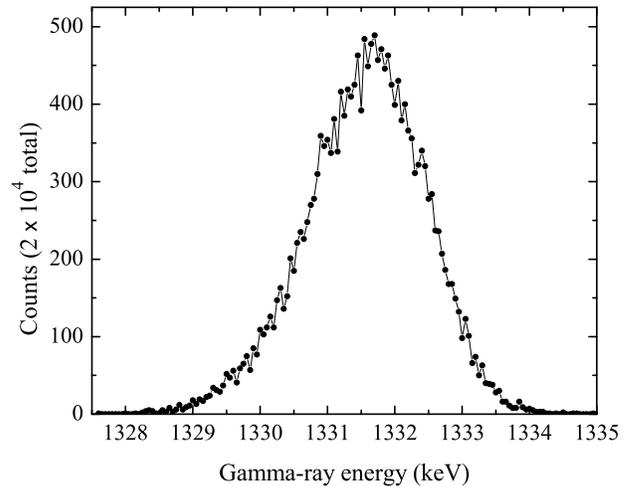}%
\caption{Peak corresponding to the 1332 keV $^{60}$Co spectral line obtained
by the n-type detector model for a neutron fluence of 10$^{9}$ cm$^{-2}$. This
model detector has the same dimensions and bias potential as the n-type
detector studied by Pehl \textit{et al.} \cite{r1}.}%
\label{fig3}%
\end{figure}

\begin{figure}[ptb]%
\centering
\includegraphics[
height=2.7259in,
width=3.2915in
]%
{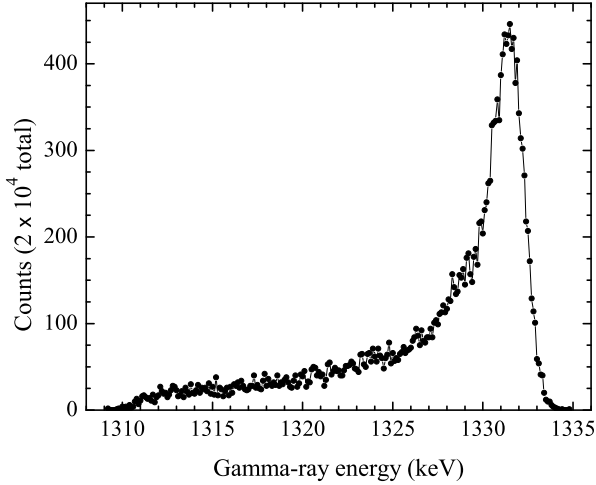}%
\caption{Peak corresponding to the 1332 keV $^{60}$Co spectral line obtained
by the n-type detector model for a neutron fluence of 10$^{10}$ cm$^{-2}$.
This model detector has the same dimensions and bias potential as the n-type
detector studied by Pehl \textit{et al.} \cite{r1}.}%
\label{fig4}%
\end{figure}

As a check (since in fact the energy resolution is very sensitive to the value
of $A_{h}$), the model gives the FWHM resolution of $64$ keV after a neutron
fluence of 10$^{10}$ cm$^{-2}$ (as shown in Fig. \ref{fig2}),%
%TCIMACRO{\FRAME{ftbpFU}{3.2915in}{2.7259in}{0pt}{\Qcb{Peak corresponding to
%the 1332 keV $^{60}$Co spectral line obtained by the p-type detector model for
%a neutron fluence of 10$^{10}$ cm$^{-2}$. This model detector has the same
%dimensions and bias potential as the p-type detector studied by Pehl
%\QTR{it}{et al.} \cite{r1}.}}{\Qlb{fig2}}{vansi2.eps}%
%{\special{ language "Scientific Word";  type "GRAPHIC";
%maintain-aspect-ratio TRUE;  display "USEDEF";  valid_file "F";
%width 3.2915in;  height 2.7259in;  depth 0pt;  original-width 4.0534in;
%original-height 3.4238in;  cropleft "0";  croptop "1";  cropright "1";
%cropbottom "0";  filename 'vansi2.EPS';file-properties "XNPEU";}}}%
%BeginExpansion
%EndExpansion
to be compared with the experimental result of $70$ keV. Figures \ref{fig3}%
%TCIMACRO{\FRAME{ftbpFU}{3.2915in}{2.7371in}{0pt}{\Qcb{Peak corresponding to
%the 1332 keV $^{60}$Co spectral line obtained by the n-type detector model for
%a neutron fluence of 10$^{9}$ cm$^{-2}$. This model detector has the same
%dimensions and bias potential as the n-type detector studied by Pehl
%\QTR{it}{et al.} \cite{r1}.}}{\Qlb{fig3}}{vansi3.eps}%
%{\special{ language "Scientific Word";  type "GRAPHIC";
%maintain-aspect-ratio TRUE;  display "USEDEF";  valid_file "F";
%width 3.2915in;  height 2.7371in;  depth 0pt;  original-width 4.0534in;
%original-height 3.4238in;  cropleft "0";  croptop "1";  cropright "1";
%cropbottom "0";  filename 'vansi3.EPS';file-properties "XNPEU";}}}%
%BeginExpansion
%EndExpansion
and \ref{fig4}%
%TCIMACRO{\FRAME{ftbpFU}{3.2915in}{2.7259in}{0pt}{\Qcb{Peak corresponding to
%the 1332 keV $^{60}$Co spectral line obtained by the n-type detector model for
%a neutron fluence of 10$^{10}$ cm$^{-2}$. This model detector has the same
%dimensions and bias potential as the n-type detector studied by Pehl
%\QTR{it}{et al.} \cite{r1}.}}{\Qlb{fig4}}{vansi4.eps}%
%{\special{ language "Scientific Word";  type "GRAPHIC";
%maintain-aspect-ratio TRUE;  display "USEDEF";  valid_file "F";
%width 3.2915in;  height 2.7259in;  depth 0pt;  original-width 4.0534in;
%original-height 3.4238in;  cropleft "0";  croptop "1";  cropright "1";
%cropbottom "0";  filename 'vansi4.EPS';file-properties "XNPEU";}}}%
%BeginExpansion
%EndExpansion
show the corresponding model results for the n-type detector
(model/experimental FWHM resolutions of $2.1/1.95$ keV and $2.7/2.7$ keV, respectively).

\begin{figure}[ptb]%
\centering
\includegraphics[
height=2.7371in,
width=3.2915in
]%
{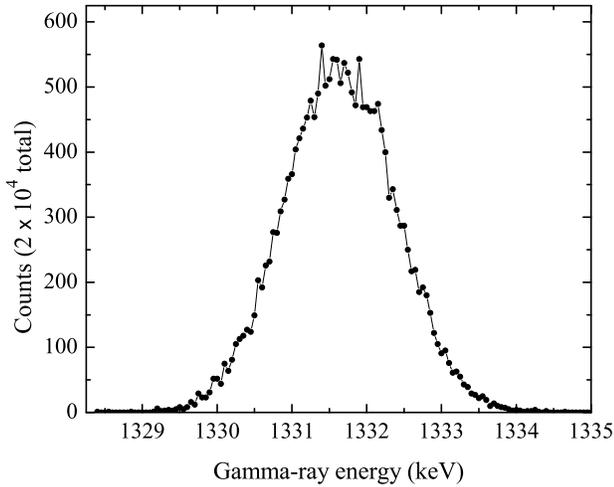}%
\caption{Peak corresponding to the 1332 keV $^{60}$Co spectral line obtained
by the p-type detector model for a neutron fluence of 10$^{8}$ cm$^{-2}$. This
model detector has the same dimensions and bias potential as the p-type
detector studied by Pehl \textit{et al.} \cite{r1}.}%
\label{fig5}%
\end{figure}

\begin{figure}[ptb]%
\centering
\includegraphics[
height=2.7371in,
width=3.2915in
]%
{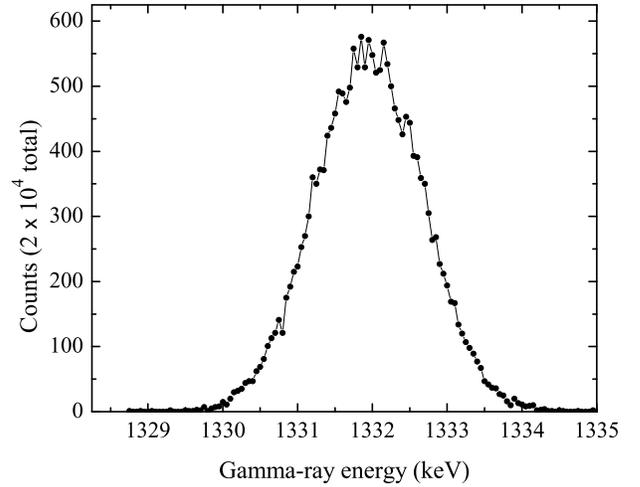}%
\caption{Peak corresponding to the 1332 keV $^{60}$Co spectral line obtained
by the n-type detector model for a neutron fluence of 10$^{8}$ cm$^{-2}$. This
model detector has the same dimensions and bias potential as the n-type
detector studied by Pehl \textit{et al.} \cite{r1}.}%
\label{fig6}%
\end{figure}

Figures \ref{fig5}%
%TCIMACRO{\FRAME{ftbpFU}{3.2915in}{2.7371in}{0pt}{\Qcb{Peak corresponding to
%the 1332 keV $^{60}$Co spectral line obtained by the p-type detector model for
%a neutron fluence of 10$^{8}$ cm$^{-2}$. This model detector has the same
%dimensions and bias potential as the p-type detector studied by Pehl
%\QTR{it}{et al.} \cite{r1}.}}{\Qlb{fig5}}{vansi5.eps}%
%{\special{ language "Scientific Word";  type "GRAPHIC";
%maintain-aspect-ratio TRUE;  display "USEDEF";  valid_file "F";
%width 3.2915in;  height 2.7371in;  depth 0pt;  original-width 4.0534in;
%original-height 3.4238in;  cropleft "0";  croptop "1";  cropright "1";
%cropbottom "0";  filename 'vansi5.EPS';file-properties "XNPEU";}}}%
%BeginExpansion
%EndExpansion
and \ref{fig6}%
%TCIMACRO{\FRAME{ftbpFU}{3.2915in}{2.7371in}{0pt}{\Qcb{Peak corresponding to
%the 1332 keV $^{60}$Co spectral line obtained by the n-type detector model for
%a neutron fluence of 10$^{8}$ cm$^{-2}$. This model detector has the same
%dimensions and bias potential as the n-type detector studied by Pehl
%\QTR{it}{et al.} \cite{r1}.}}{\Qlb{fig6}}{vansi6.eps}%
%{\special{ language "Scientific Word";  type "GRAPHIC";
%maintain-aspect-ratio TRUE;  display "USEDEF";  valid_file "F";
%width 3.2915in;  height 2.7371in;  depth 0pt;  original-width 4.0534in;
%original-height 3.4238in;  cropleft "0";  croptop "1";  cropright "1";
%cropbottom "0";  filename 'vansi6.EPS';file-properties "XNPEU";}}}%
%BeginExpansion
%EndExpansion
show the model results for the p-type and n-type detectors, respectively,
\ after a neutron fluence of 10$^{8}$ cm$^{-2}$ (model/experimental FWHM
resolutions of $1.85/2.1$ keV and $1.80/1.8$ keV, respectively). For this low
fluence the model FWHM resolutions are nearly identical for the two detectors;
instead the effect of hole trapping shows up in the magnitude of the shift of
the peak centroid away from 1332 keV.

\section{Application to the INL Micro-Detective}

\begin{figure}[ptb]%
\centering
\includegraphics[
height=2.7614in,
width=3.1522in
]%
{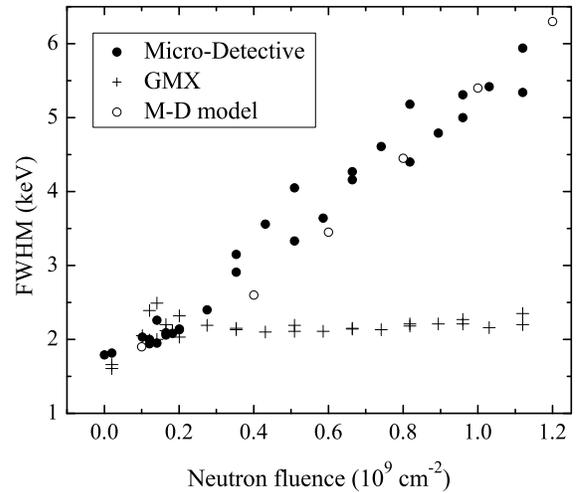}%
\caption{Measured FWHM resolution of the 1332 keV $^{60}$Co peak obtained by
the Micro-Detective and GMX detectors after various neutron fluences (where
two values are shown for the same fluence, the higher/lower value was measured
before/after a thermal cycle). The calculated values obtained by the
Micro-Detective model are indicated by the open circles.}%
\label{fig7}%
\end{figure}

We exposed an ORTEC Micro-Detective (p-type) detector to a neutron fluence
(from a $^{252}$Cf source) up to 10$^{9}$ cm$^{-2}$. This exercise was
intended to determine the neutron fluence at which the energy resolution of
the Micro-Detective was too degraded to allow its use in place of our GMX
(n-type) detector. Figure \ref{fig7}%
%TCIMACRO{\FRAME{ftbpFU}{3.1522in}{2.7614in}{0pt}{\Qcb{Measured FWHM resolution
%of the 1332 keV $^{60}$Co peak obtained by the Micro-Detective and GMX
%detectors after various neutron fluences (where two values are shown for the
%same fluence, the higher/lower value was measured before/after a thermal
%cycle). The calculated values obtained by the Micro-Detective model are
%indicated by the open circles.}}{\Qlb{fig7}}{vansi7.eps}%
%{\special{ language "Scientific Word";  type "GRAPHIC";
%maintain-aspect-ratio TRUE;  display "USEDEF";  valid_file "F";
%width 3.1522in;  height 2.7614in;  depth 0pt;  original-width 3.9003in;
%original-height 3.4636in;  cropleft "0";  croptop "1";  cropright "1";
%cropbottom "0";  filename 'vansi7.EPS';file-properties "XNPEU";}}}%
%BeginExpansion
%EndExpansion
shows values of the FWHM resolution of the 1332 keV $^{60}$Co peak obtained by
the Micro-Detective and GMX detectors at various fluences (where two values
are provided for the same fluence, the higher/lower value was measured
before/after a thermal cycle). These results provide an opportunity to test
the phenomenological model.

The input to the Micro-Detective model is the following: inner radius
$R_{0}=4.5$ mm; outer radius $R_{1}=25$ mm; crystal length $L=30$ mm; bias
potential $\phi\left(  R_{1}\right)  =3.0$ kV. As the dopant density is not
provided, the typical value $\rho_{\text{A}}=10^{10}$ cm$^{-3}$ is used. The
calculated values of the FWHM resolution, indicated by the open circles in
Fig. \ref{fig7}, compare well with the trend of measured values. Note that the
energy resolution of the Micro-Detective after a neutron fluence of 10$^{9}$
cm$^{-2}$ is better than that of the p-type detector studied by Pehl
\textit{et al.} ($5.4$ keV versus $6$ keV) despite its significantly larger
size (diameter $50$ mm versus $42$ mm), due to its larger bias potential ($3$
kV versus $1.6$ kV) which, by producing a stronger electric field across the
crystal, reduces the hole trap cross-section.

The input to the GMX (n-type) detector model is the following: inner radius
$R_{0}=5.6$ mm; outer radius $R_{1}=31.4$ mm; crystal length $L=70.3$ mm; bias
potential $\phi\left(  R_{0}\right)  =5.0$ kV. As the dopant density is not
provided, the typical value $\rho_{\text{D}}=10^{10}$ cm$^{-3}$ is used. The
calculated FWHM resolutions of the 1332 keV peak at neutron fluences of
10$^{8}$ cm$^{-2}$ and 10$^{9}$ cm$^{-2}$ are $1.7$ keV and $2.0$ keV,
respectively, in good agreement with the measured values in Fig. \ref{fig7}.

\section{Concluding remarks}

The main attributes of this model of neutron-damaged coaxial HPGe detectors
are (i) the use of induced charge at the electrodes to determine the
contribution of an electron-hole pair to the measured gamma-ray energy, and
(ii) the use of stochastic methods to simulate what are, in fact, stochastic
processes. The data of R. H. Pehl \textit{et al.} \cite{r1} provided a value
for the dimensionless parameter $A_{h}$, related to the probability for
immediate trapping of a mobile hole, needed to complete the phenomenological
model. As the model is, for the most part, one-dimensional, it is easy to
implement in a computer code. However, by ignoring the detector
\textquotedblleft cap\textquotedblright\ (where the applied electric field is
not purely radial), this model is not well suited for application to
low-energy gamma-rays which may be substantially stopped in that volume.

Some observations: (i) The n-type detectors maintain good energy resolution to
neutron fluences of at least 10$^{9}$ cm$^{-2}$. The noticeable effect of
neutron damage is to shift the peak centroid to a lower energy. (ii) The peak
shapes after a high neutron fluence of 10$^{10}$ cm$^{-2}$ are very different
for p-type and n-type detectors. This is due not only to more hole trapping in
a p-type detector, but also to the fact that an electron-hole pair with a
trapped hole near the outer contact induces a smaller charge $\left\vert
Q\right\vert $ at the two electrodes of a p-type detector than at the
electrodes of a same-sized n-type detector (see section II for a more precise
discussion of this effect). Thus those gamma-rays that interact closer to the
outer contact (which is to say, most of the gamma-rays) are more likely to
register as \textquotedblleft counts\textquotedblright\ at the low/high end of
the energy spectrum in the case of p-type/n-type coaxial detectors.


\begin{thebibliography}{9}                                                                                                %


\bibitem {r1}R. H. Pehl, N. W. Madden, J. H. Elliott, T. W. Raudorf, R. C.
Trammell, and L. S. Darken, Jr., \textquotedblleft Radiation damage resistance
of reverse electrode Ge coaxial detectors,\textquotedblright\ \textit{IEEE
Trans. Nucl. Sci.}, vol. NS-26, no. 1, pp. 321--323, Feb. 1979.

\bibitem {r1b}J. D. Jackson, \textit{Classical Electrodynamics}. New York:
Wiley, 1962.

\bibitem {r1e}G. F. Knoll, \textit{Radiation Detection and Measurement}, 3rd
ed. New York: Wiley, 2000.

\bibitem {r2}G. E. P. Box and M. E. Muller, \textquotedblleft A note on the
generation of random normal deviates,\textquotedblright\ \textit{Ann. Math.
Statist.}, vol. 29, no. 2, pp. 610--611, 1958.

\bibitem {r3}L. S. Darken, Jr., R. C. Trammell, T. W. Raudorf, R. H. Pehl, and
J. H. Elliott, \textquotedblleft Mechanism for fast neutron damage of Ge(HP)
detectors,\textquotedblright\ \textit{Nucl. Instrum. Methods}, vol. 171, pp.
49--59, 1980.

\bibitem {r4}T. W. Raudorf and R. H. Pehl, \textquotedblleft Effect of charge
carrier trapping on germanium coaxial detector line shapes,\textquotedblright%
\ \textit{Nucl. Instrum. Methods Phys. Res. A}, vol. A255, pp. 538--551, 1987.

\bibitem {r4e}L. S. Darken, \textquotedblleft Role of disordered regions in
fast-neutron damage of HPGe detectors,\textquotedblright\ \textit{Nucl.
Instrum. Methods Phys. Res.}, vol. B74, pp. 523--526, 1993.

\bibitem {r5}L. S. Darken, Jr., R. C. Trammell, T. W. Raudorf, and R. H. Pehl,
\textquotedblleft Neutron damage in Ge(HP) coaxial
detectors,\textquotedblright\ \textit{IEEE Trans. Nucl. Sci.}, vol. NS-28, no.
1, pp. 572--578, 1981.

\bibitem {r6}Th. Kr\"{o}ll and D. Bazzacco, \textquotedblleft Simulation and
analysis of pulse shapes from highly segmented HPGe detectors for the $\gamma
$-ray tracking array MARS,\textquotedblright\ \textit{Nucl. Instrum. Methods
Phys. Res. A}, vol. 463, pp. 227--249, 2001.
\end{thebibliography}
\end{document}